# Impact of Interdigitated Electrodes design on the Low Frequency and Random Telegraph Noise of Single-Layer Graphene Micro Ribbons

**Authors:** Georgia Samara [a,b] *, Christoforos Theodorou [c], Alexandros Mavropoulis [a], Nikolaos Vasileiadis [a], Konstantinos Papagelis [b], Panagiotis Dimitrakis [a,d] *

[a] Institute of Nanoscience and Nanotechnology, NCSR "Demokritos", Ag. Paraskevi, Greece

[b] Department of Physics, Aristotle University of Thessaloniki, Thessaloniki, Greece

[c] Univ. Grenoble Alpes, Univ. Savoie Mont Blanc, CNRS, Grenoble INP, CROMA, Grenoble, France

[d] Institute of Quantum Computing and Quantum Technology, NCSR "Demokritos", Ag. Paraskevi, Greece

## Abstract

High performance devices consisting of interdigitated electrodes (IDEs) on top of single-layer graphene (SLG) are candidates with favorable prospects for sensing applications. Graphene micro ribbons (GMRs) of various widths and IDE design geometries were fabricated and experimentally examined regarding their low-frequency noise (LFN) behavior. Measurements revealed a 1/f behavior and different kinds of trap activity behind it, which were studied through the analysis of random telegraph noise (RTN) signals. Our investigation suggests that adjusting the geometrical characteristics of either the GMR width or the IDE topology can significantly influence the signal-to-noise ratio (SNR) of SLG-based electronics. On the bright side, the results of our study can provide useful guidelines for fabrication decisions to maximize the SNR.

* Corresponding authors
Email address: g.samara@inn.demokritos.gr
　　　　　　　p.dimitrakis@qi.demokritos.gr

# I. Introduction

Building on the impressive development of classic semiconductors, such as Si, over the past decade, two-dimensional (2D) materials have captured scientific interest due to the excellent merging of downscaled dimensions and enhanced material properties in order to enable applications in the "More than Moore" technology direction. Boehm et al. [1] were pioneers in the investigation and isolation of graphene in 1961, as well as in its definition through the graphite lattice in 1994. In 2004, graphene was the cornerstone for in-depth research, initially in atomically thin carbon sheets through its isolation utilizing the exfoliation method [2] and then in other materials with similar outstanding properties on the atomic scale. SLG is a zero-bandgap semiconductor with a structure resembling a honeycomb planar lattice of carbon atoms coupled via $\pi$-$\pi$* bonds [3], [4], [5]. Moreover, it has been characterized as a supreme material due to its exceptionally high electron mobility at room temperature (highest theoretical limit at 200,000 $cm^2V^{-1}s^{-1}$), absorbance of white light equal to 2.3%, thermal conductivity ranging from 4.84 to 5.30 x $10^3$ $Wm^{-1}K^{-1}$[6], breaking strength 200 times higher than that of steel, Young's modulus of 1 TPa and low fabrication cost [7], [8], [9], [10]. Thus, a lot of research effort was directed towards SLG because it has potential for a wide spectrum on electronic and sensing applications [11].Electrical noise measurements can serve as a diagnostic tool for thermal, shot, burst, generation-recombination, $1/f$ and $1/f^2$ noise, providing insights into the relationship between the noise in a device and its reliability, or its intrinsic properties such as interface trap density [12]. The first statements on LFN measured in graphene devices were addressed in 2007 and 2008, respectively [13], [14]. A comprehensive study of LFN in SLG devices was performed in 2013 by Balandin's group [15], demonstrating a $1/f$ spectral dependence. More recently, in 2022, Nah et al. investigated various IDE topologies of functionalized SLG sheets and examined their resistive properties for sensing applications [16]. However, the noise behavior of such different IDE topologies has not yet been examined. Recently, we reported preliminary results of low-frequency noise measurements in similar GMR-IDE structures [17]. In the present research work, we extend our previous studies and provide a comprehensive analysis of the measured noise signals. Specifically, different IDE configurations on top of different pristine SLG micro-ribbons were manufactured, and the samples were thoroughly examined through Current-Voltage (I-V) characteristics, LFN, and RTN measurements. The possibility of a direct dependence between the measured electronic properties and the geometrical characteristics of the IDEs is investigated. The aim of this study is to use these topologies as sensing devices. A key aspect is to investigate the trade-off between GMR width and the distance among IDEs in order to maximize the signal-to-noise ratio (SNR).

## II. Materials and Methods

### A. Device fabrication process

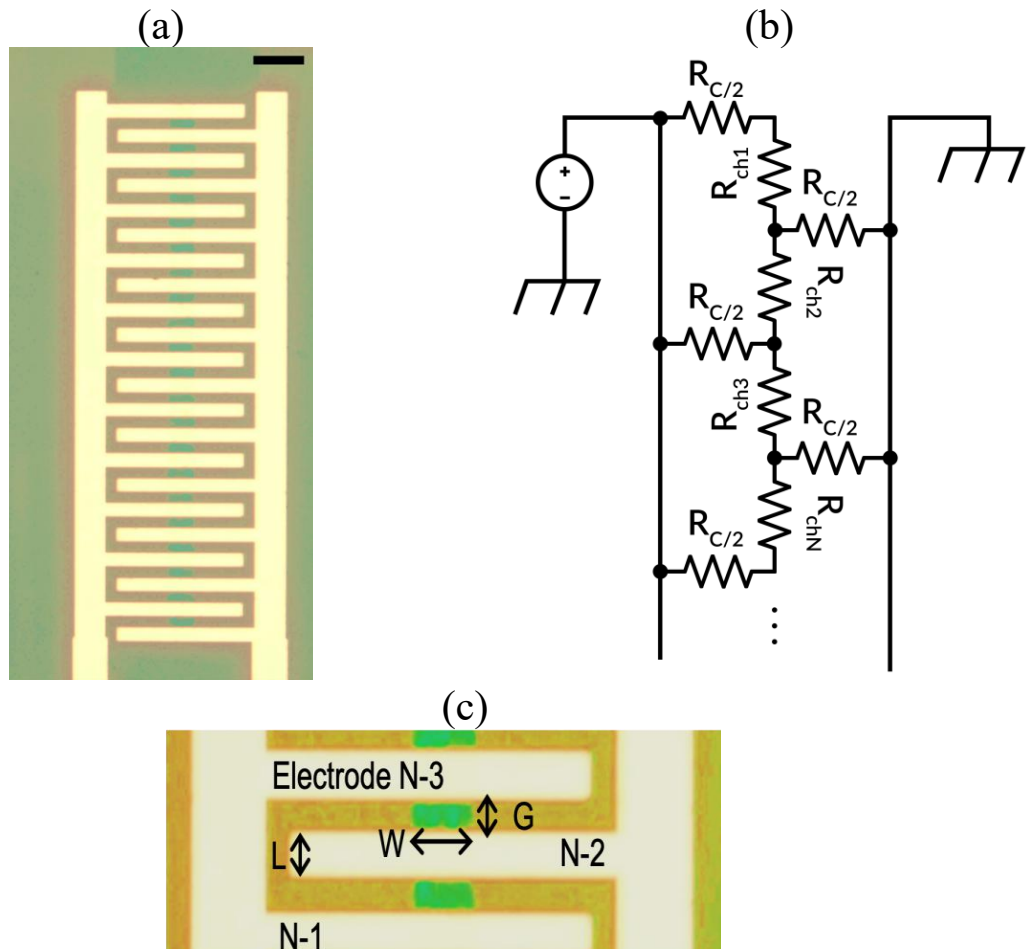

Figure 1: (a) Optical microscope image and (b) electrical equivalent circuit of a typical IDE-GMR device. Inset: the scale bar corresponds to 100 µm. (c) Definition of the geometric parameters of the IDE-GMR architecture. Here, G= 25 µm, W= 50 µm, L= 25 µm.

Chemical vapor deposition (CVD) technique was employed for SLG growth on Cu foil and coated with poly(methyl methacrylate) (PMMA). Cu foil was etched in an Ammonium Persulfate solution, and then a wet-transfer process was utilized to transfer SLG/PMMA on top of a 300 nm $SiO_2$/Si substrate. PMMA was removed by UV exposure (325 nm) and then developed in methyl isobutyl ketone (MIBK) : isopropanol (IPA) (3:1) solution. The last step before graphene patterning was furnace annealing in hydrogen, contributing to PMMA residual removal. Subsequently, spin coating, electron beam lithography (EBL), and Oxygen plasma (dry) etching techniques were applied in order to pattern the GMR strips (Fig. 1). Different topologies of IDEs, i.e. IDE with various electrode interspace distances or gaps G (8, 15, and 25 μm) on top of GMRs having different widths W (50, 100, and 200 μm), were fabricated by the Aluminum lift-off process. A total number of nine devices were fabricated following the combinations of the different gaps and widths. A typical optical microscope image of the final IDEs (11 electrode pairs with interspace distance 25 μm) and GMR (width of 50 μm) topology is presented in Fig. 1(a). The corresponding electrical equivalent circuit is depicted in Fig. 1(b), in which $R_{ch}$ and $R_{c/2}$ denote the SLG and contact resistances, respectively.

## B. Electrical Characterization Methods

An HP4155A Semiconductor Parameter Analyzer integrated with a wafer prober was used to perform Current-Voltage (I-V) measurements. To conduct RTN measurements, the instrumentation setup illustrated in Fig. 2 was used, wherein the device under test (DUT) was interfaced with a low-noise current amplifier (SR570), a Keysight B2902A source-measure unit, and a Keysight DSO7104A digital oscilloscope.

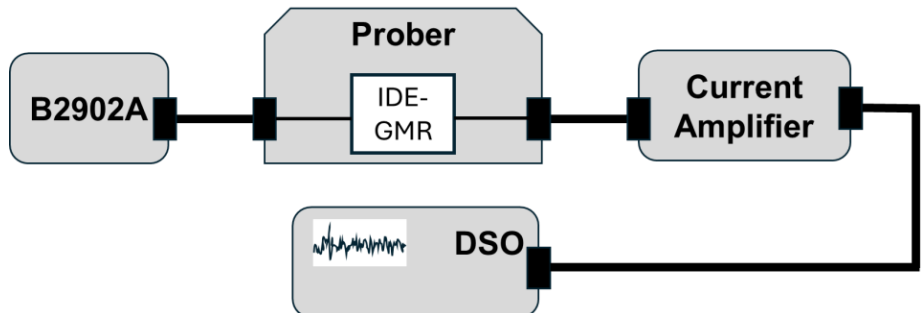

Figure 2: Final setup for RTN measurements.

# III. Results and Discussion

## A. Current-Voltage Measurements

All the fabricated IDE-GMR devices were tested through I-V measurements, which were performed by applying logarithmic current sweeps in the range of 1 to 100 µA and measuring the corresponding voltage. It should be emphasized that no bias was applied to the backside (floating) of the sample. The experimental results for different G gaps are shown in Fig. 3, with the GMR width (W) considered as a parameter. The work function difference between GMR and aluminum (Al) electrodes is of great importance, as it governs the Ohmic behavior of the contacts. In accordance with measurements in Fig. 3, the experimental data within the range 1-20 µA fit excellently to the relation $ln(V) = ln(R) + n \cdot ln(I)$, with $0.98 \leq n \leq 1$, revealing Ohmic conduction. For higher current values, 20-100 µA, the experimental data deviate from linearity, likely due to the self-heating effect of GMR and/or conduction through grain boundaries.

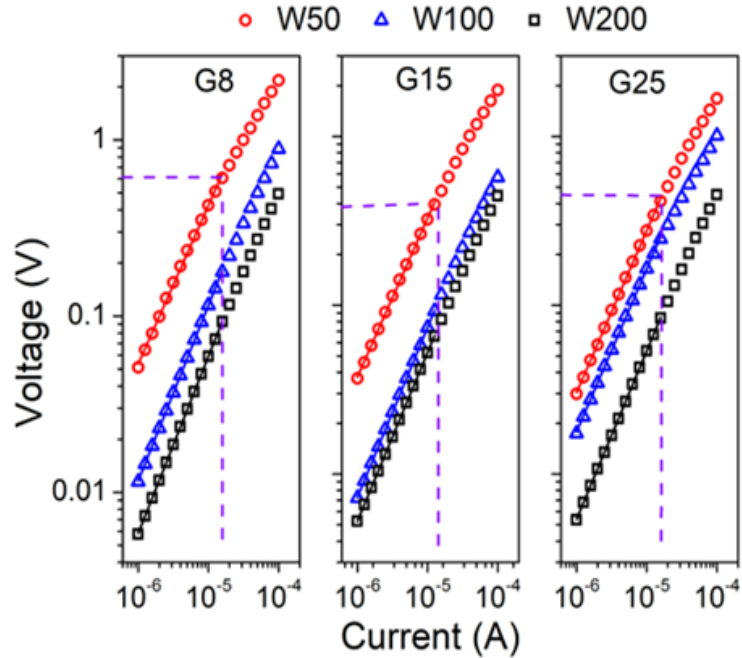

Figure 3: The Current-Voltage characteristics of IDE-GMR devices with different graphene ribbon widths (W) and electrode interspace distances (G).

This effect becomes stronger as *W* decreases. Furthermore, as expected, devices with higher W/G ratios exhibit lower resistance. Indeed, the resistance *R* of a thin film material with resistivity *ρ*, length *L*, width *W,* and thickness *t* is given by the well-known formula *R=ρL/Wt*. In our case L= G and for the same GMR, resistivity and thickness can be considered constant; hence, *R=R$_{sheet}$·[G/W(N$_{IDE}$-1]*, where *R*$_{sheet}$ is the sheet resistance of the GMR in Ω/□. The resistance values calculated from the measured data after linear fitting, as shown in Fig. 3, are presented in Table 1. The estimated error in resistance is lower than 1‰. Notably, the resistance values listed in Table 1 are relatively high compared to the values of pristine SLG reported in the literature [18]. This is mainly attributed to the high contact resistance of Al on SLG [19], [20].

Where $N_{IDE}$ is the number of interdigitated electrodes.

Table 1. The extracted resistance values after linear fitting on data shown in Fig. 3.

| G(µm) | W50 (W=50µm) | | | W100 (W=100µm) | | | W200 (W=200µm) | | |
|---|---|---|---|---|---|---|---|---|---|
| | W($N_{IDE}$-1)/G | R (kΩ) | $R_{sheet}$ (kΩ/□) | W($N_{IDE}$-1)/G | R (kΩ) | $R_{sheet}$ (kΩ/□) | W($N_{IDE}$-1)/G | R(kΩ) | $R_{sheet}$ (kΩ/□) |
| 8 | 194 | 38 | 7372 | 388 | 11 | 4268 | 775 | 6 | 4650 |
| 15 | 117 | 29 | 3393 | 233 | 7 | 1631 | 467 | 5 | 2335 |
| 25 | 42 | 25 | 1050 | 84 | 13 | 1092 | 168 | 5 | 840 |

## B. Low Frequency Noise characteristics and 1/*f* behavior

LFN measurements were performed for all different topologies by applying a bias voltage between the electrodes in the range 0.1 V ≤ *V* ≤ 0.6 V, with a floating bottom gate. For each bias voltage, the current was recorded for 100 s with a sampling frequency of 80 kHz. Following, the *Power Spectral Density* (PSD) was obtained by applying the *Fast Fourier Transform* (FFT) (Welch method) to all *I-t* recordings (Fig. 4(a)).

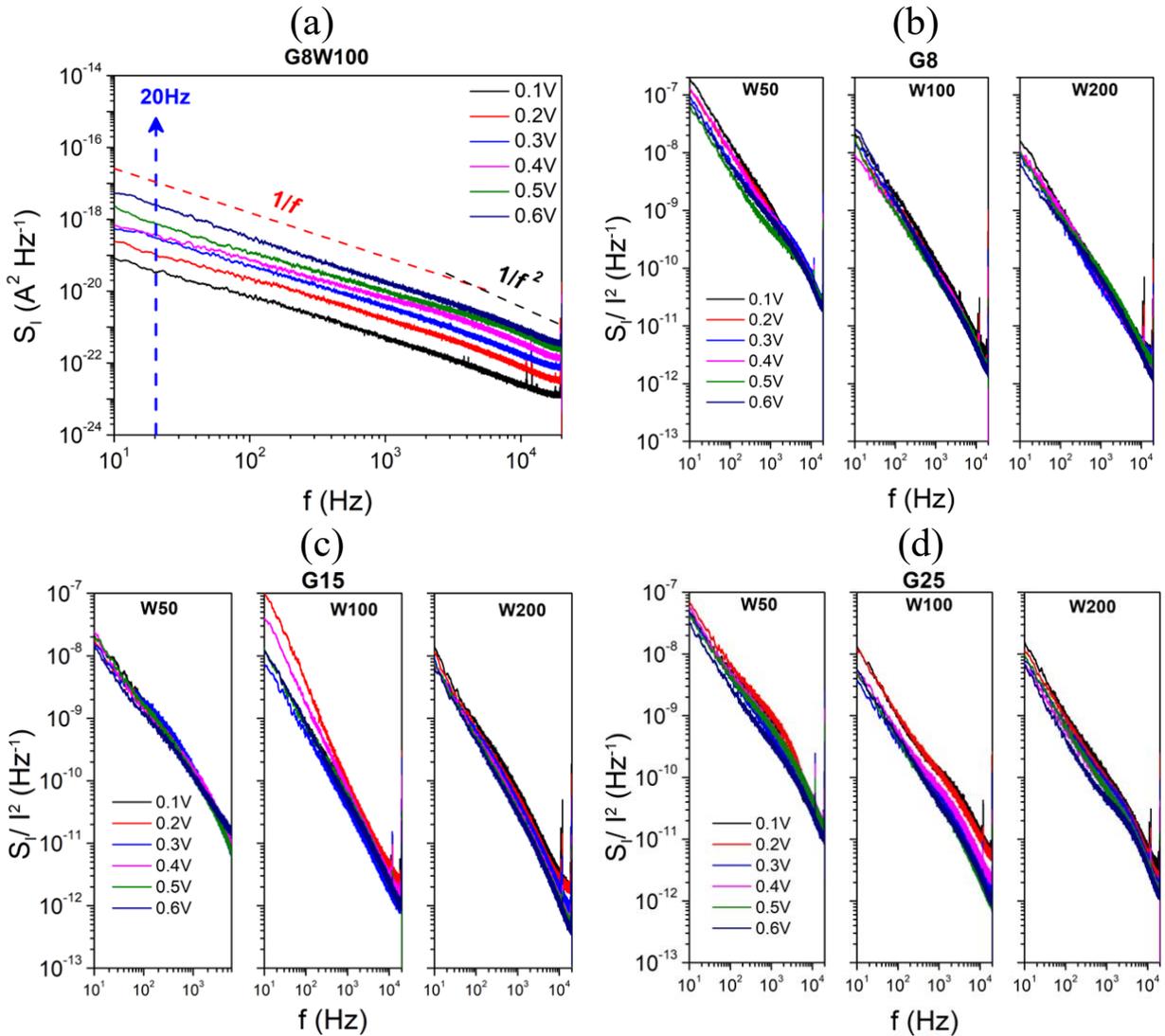

Figure 4: (a) Typical PSD spectra for the IDE-GMR with G = 8 μm and W = 100 μm, illustrating the corresponding theoretical *1/f* and *1/f²* dependencies. (b), (c), and (d) Normalized PSD spectra $S_I/I^2$ versus *f* for all tested devices. I-t measurements were scrutinized for different constant biases ranging from 0.1 to 0.6 V.

The PSD spectra for all applied voltages are shown in Fig. 4(a). It is evident that the $1/f^\gamma$ dependence of the PSD is observed at any bias voltage with $1 \leq \gamma \leq 2$. Subsequently, Figs. 4 (b), (c), and (d) depict the log-log plots of the normalized PSD ($S_I / I^2$) vs. $f$ in the range of 10 Hz to 20 kHz, organized and studied in triplets with fixed electrode interspace distance. All spectra clearly exhibit the $1/f^\gamma$ dependence. Specifically, under different bias voltages, PSD plots exhibited various slopes, γ, on log-log scale, from γ=0.5 to γ=2, depending on the measured sample and bias condition, which indicates the presence of Lorentzian spectrum contributions, commonly related to isolated traps. According to the I-V curves shown in Fig. 3, the application of voltages higher than 0.6 V affects the ohmic behavior of the GMR and hence the noise measurements. To discover the origin of $1/f$ noise, all measured $S_I$ values in a defined range of constant applied bias (0.1 V ≤ $V$ ≤ 0.5 V) across various topologies, were extracted at a specific frequency value set to 20 Hz. Furthermore, plots of $S_I^{1/2}$ (at 20Hz) vs voltage were generated and presented in Fig. 5(a) indicating that the $1/f$ noise amplitude is linearly dependent on the applied voltage. The observed dependence indicates that the measured noise originates from GMR's conductance fluctuations due to traps (variation of γ with frequency) and not from contact-related defects, as demonstrated below. Almost identical trends were observed in all studied triplets. Following, based on Equation (1), the normalized square root resistance noise, $S_R^{1/2}/R^2$, term was calculated from the data depicted in Fig. 5(a).

$$\frac{S_I}{I^2} = \frac{S_R}{R^2} \Rightarrow \frac{\sqrt{S_I}}{I} = \frac{\sqrt{S_R}}{R} \Rightarrow \frac{\sqrt{S_I}}{V} = \frac{\sqrt{S_R}}{R^2} \Rightarrow \sqrt{S_I} = \frac{\sqrt{S_R}}{R^2} V \quad (1)$$

where $R=V/I$ is the total resistance of the tested device (see Table 1), $S_I$ and $S_R$ are the current and resistance noise, respectively. Under severe impact of contact resistance $R_C$, the measured LFN spectra are described by the following relation [21], [22] :

$$\left(\frac{S_I}{I^2}\right)_M = \left(\frac{S_I}{I^2}\right)_G + \left(\frac{I}{V}\right)^2 S_{Rc} \quad (1a)$$

Where the subscripts M and G stand for measured and Graphene normalized PSD respectively. Indeed, according to data shown in Table 1, these two devices exhibit the highest contact resistances. Since Resistance ∝ 1/area, it is expected that resistance noise depends on the reciprocal value of the contact area.

According to McWhorter and Van der Ziel [23], the carrier number fluctuations PSD, $S_N$, due to trapping/de-trapping in slow oxide traps near the interface can be described by:

$$\frac{S_N}{N^2} = \frac{S_{Nt}}{N^2} = \frac{A_G N_t kT}{N^2 f} \quad (2)$$

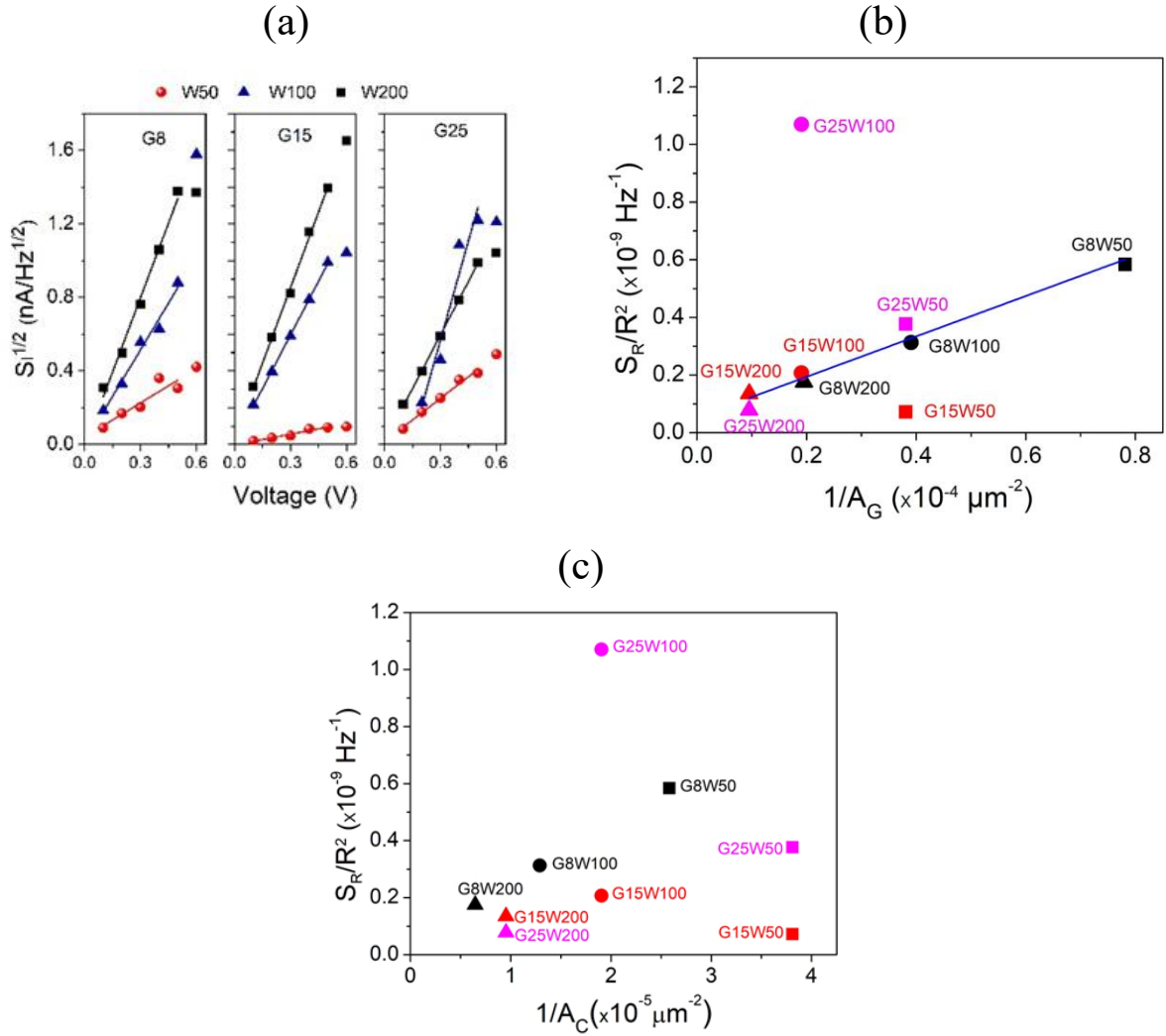

Figure 5: (a) $S_I^{1/2}$ vs. Voltage at 20 Hz, (b) $S_R^{1/2}/R^2$ vs. $1/A_G$ and (c) $S_R^{1/2}/R^2$ vs. $1/A_C$ for all tested IDE-GMR devices, where $A_G$ and $A_C$ denote the areas of free graphene and contact regions, respectively. A linear fit to the experimental data is shown where applicable

Where N is the total number of carriers in the conductive volume, $A_G$ is the active graphene surface area given by $A_G = G \cdot W \cdot (N_{IDE}-1)$ and $N_t$ is the trap density in $cm^{-2}eV^{-1}$. Assuming a constant GMR mobility $\mu$ under the applied voltage bias V, and that the current at low V is equal to:

$$I = \frac{W(N_{IDE}-1)}{G} qn\mu V = \frac{W(N_{IDE}-1)}{G} q \frac{N}{W(N_{IDE}-1)G} \mu V = q \frac{N}{G^2} \mu V \quad (3)$$

where n the GMR carrier density in $cm^{-2}$, one can write:

$$\frac{S_I}{I^2} = \frac{S_N \left(\frac{q\mu V}{G^2}\right)^2}{N^2 \left(\frac{q\mu V}{G^2}\right)^2} = \frac{S_N}{N^2} \quad (4)$$

Combining (2) and (4) we obtain:

$$\frac{S_I}{I^2} = \frac{A_G G N_t kT}{N^2 f} = \frac{A_G N_t kT}{n^2 A_G^2 f} = \frac{N_t kT}{n^2 A_G f} \quad (5)$$

which reveals a potential scaling of the normalized PSD with the reciprocal of area, assuming a constant carrier density, n. Based on (5), we plotted the normalized resistance noise, $S_R/R^2$, as a function of $1/A_G$, where $N_{IDE}$ is the number of the IDE as defined in Fig. 5(b), to verify whether any linear dependence is observed. Each combination of symbol and color corresponds to a device with a specific G and W topology. Indeed, a linear fit was successfully applied to these datasets, with the exception of two outlier devices (G15W50 and G25W100) which deviate from the linear regression. This may be attributed to the presence of wrinkles or abnormalities in the graphene sheet as well as to the contribution of contact resistance noise, $S_{RC}$. Under severe influence of contact resistance $R_C$, the measured LFN spectra would be dominated by $S_{RC}$. In that case, according to (5), the normalized PSD is expected to scale with $1/A_C$, where $A_C = W \cdot (N_{IDE}-1)/L$.

To examine this hypothesis, the $S_R/R^2$ is plotted as a function of the inverse contact area $1/A_C$, in Fig. 5(c). Evidently, there is no clear correlation or linear dependence between these data, not even for the two outliers. In summary, Fig. 5 indicates that LFN results mainly from SLG conductance fluctuations rather than from contact region noise. In addition, Fig. 5(b) also provides insights into the noise performance with respect to IDE topology, indicating that an augmentation in the active GMR area results in a decreased SNR value. Thus, geometrical characteristics of the devices have a critical impact on LFN and consequently on device performance.

C. Characterization of Random Telegraph Noise

Apart from frequency-domain noise analysis for each device, time-domain noise analysis also provides significant information. In Fig. 6, a typical example of current amplitude distribution is shown for the G8W100 device under six different constant applied voltages. The experimental current distributions fit perfectly with single- and-multi-Gaussian distributions (0.4 and 0.5 V), revealing the presence of RTN signals.

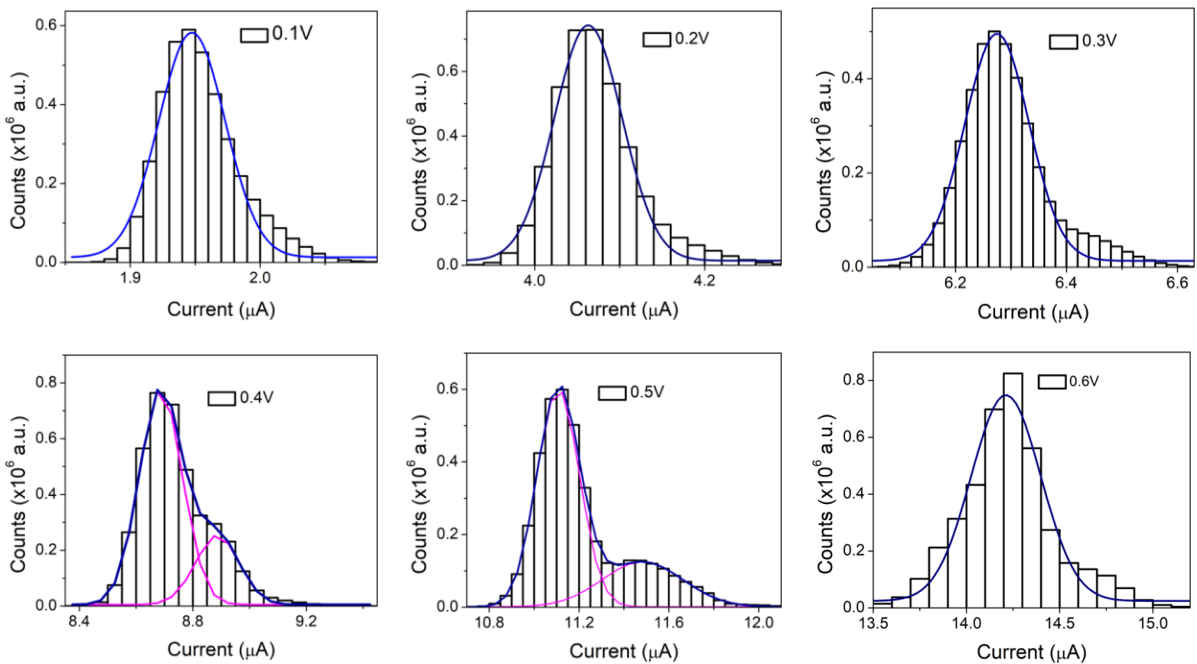

Figure 6: Histograms were extracted from the statistical analysis of the recorded *I-t* waveforms of the G8W100 device at various constant applied voltages. The Gaussian curve fitting approach was used in most cases, while multi-Gaussian fitting was applied to the 0.4 V and 0.5 V datasets.

Typical examples of RTN are depicted in Fig. 7. Specifically, RTN time windows are demonstrated together with the corresponding histograms with multi-Gaussian distribution. For G8W100 topology, the *I-t* waveform captured under a constant voltage bias of 0.5V, more than one switching event was observed at different time windows, as shown in Figs. 7(b), and (c) for example.

The varying number of RTN levels suggests the presence of one or more traps [24]. Specifically, Figs. 7(a), (c), and (e) present two-level RTN signals, while Figs. 7(b), (d), and (f) show three-level RTN signals.

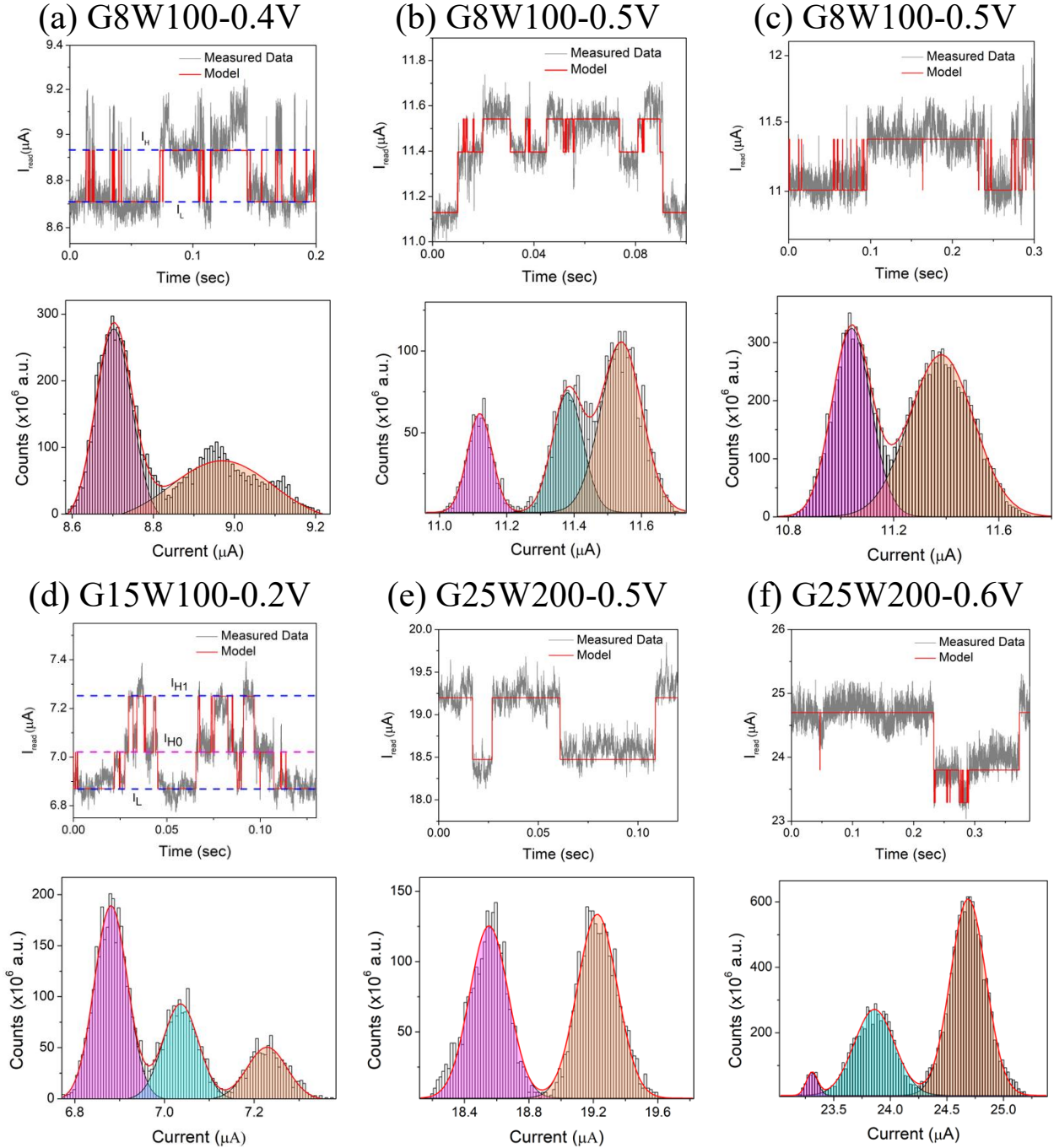

Figure 7: (a)-(f) Typical illustrations of RTN signals detected in *I-t* waveforms, along with the corresponding vertical histograms plotted after statistical analysis.

Three-level RTN seems to suggest that, during a certain time window, electrons are captured and emitted by two traps, while a two-level RTN signal corresponds to the presence of a single trap. The different levels in the random telegraph signal were extracted from the multi-Gaussian fitting of the corresponding counts histogram.

This is also represented in the corresponding Weighted Time Lag Plots (WTLP) (see Supplementary). In all time-domain noise illustrations, high and low current levels of RTN noise are denoted by dotted lines (blue) and named $I_H$ and $I_L$, respectively. In three-level noise recordings, such as in Figs. 7(b), (d), and (f), there is an additional intermediate level (pink dotted line), named $I_{H0}$. The final objective of this study was to identify the origin of RTN events between traps at the channel/dielectric and/or channel/metal contact interfaces, where the channel is defined as the material through which electrons flow. In the examined devices, there are two possible interfaces: the SLG/SiO$_2$ (channel and bottom dielectric region) and the SLG/Al (contact regions). As discussed in Section B, the analysis of normalized noise resistance revealed that LFN is attributed to resistance fluctuations in the graphene layer and not to series resistance at the contacts. The $\Delta I/I_H$ (%) ratio, where $\Delta I$ is the difference between the current mean values of two sequential RTN levels, e.g., $I_H$ and $I_L$, was extracted for all detected RTN signal levels.

To distinguish whether the RTN originates from SLG/SiO$_2$ or SLG/Al, we plotted $\Delta I/I_H$ versus the inverse of the total graphene channel area, $A_G$ (Fig. 8(a)), and versus the inverse of the contact region area, $A_C$ (Fig. 8(b)). This was done because, based on (3), and considering no mobility change due to the trapping, a single electron trapping, which introduces a $\Delta N$ change equal to 1, would yield a $\Delta I/I$ equal to:

$$\frac{\Delta I}{I} = \frac{\Delta n}{n} \times \frac{1}{A_{G,C}} \quad (6)$$

For each IDE-GMR device, more than one RTN level was detected; therefore, the extracted $\Delta I/I_H$ values were categorized and studied in three regions: RTN-1, ranging from 0 to 3%; RTN-2, from 3 to 5%; and RTN-3, for values greater than 5%, depicted with black, red, and blue colors, respectively. Since the examined factor $1/A_G$ in Fig. 8(a) includes both G and W variables, it was not clear whether the traps are related to the contact region or the free graphene area. After performing linear fitting in the RTN-1 and RTN-2 regions of Fig. 8(a), Pearson's R linearity coefficient values were determined to be 0.9 and 0.79, respectively. However, in the same regions of Fig. 8(b) these values were equal to 0.38 (RTN-1) and 0.91 (RTN-2). Hence, it can be claimed that the RTN-1 (low-amplitude) traps originate from $SiO_2$/SLG interface, while RTN-2 (medium-amplitude) traps are located within the contact region, where the $SiO_2$/SLG and/or SLG/Al interfaces are involved. However, since trapping at SLG/Al would imply not only the presence of dangling bonds in the graphene layer but also much faster traps due to the high free carrier concentration in the ohmic contact, and in our case (Fig. 7) the capture/emission times are in the millisecond range, we conclude that all RTN levels are generated by $SiO_2$ interface traps in both channel and contact regions, with contact regions resulting in higher amplitudes. Interestingly, G15W50 is an outlier in all RTN regions in both illustrations. This result in conjunction with Fig. 5(b) (1/$f$ noise analysis) indicates that the SLG quality of this device is lower than the rest, which may result from the fabrication process.

It is significant to mention that, in both plots (Fig. 8(a) and (b)), the RTN-3 group does not exhibit any direct dependence between $\Delta I/I_H$ (%) and the geometrical characteristics of the device. We assess that this behavior can be attributed to non-uniformities or percolation effects in the graphene conduction area, acting as a single trap that is strong enough to significantly reduce the current in the conduction path, leading to RTN amplitudes much higher than those expected from (6).

Moreover, Fig. 8(c) was constructed to examine whether the RTN amplitudes scale with the inverse of the total measured conductance, $G_{tot} = dI/dV$, calculated from the derivatives of the I-V plots of Fig. 3. The reason is that whatever the physical source of fluctuation behind the RTN, i.e., carrier number fluctuations, induced mobility fluctuations or series resistance fluctuations- all of them are reflected in the total conductance fluctuation [25], because:

$$I = G_{tot}V \Rightarrow dI = dG_{tot}V = \frac{I}{G_{tot}} dG_{tot} \Rightarrow \frac{\Delta I}{I} = \frac{\Delta G_{tot}}{G_{tot}} \quad (7)$$

Firstly, the illustration of $\Delta I/I_H$ (%) compared to $1/G_{tot}$ suggests a direct correlation between the width of SLG and the total conductance. Even without knowing how n or μ vary between different devices or regions, the experimental results of Fig. 8(c) suggest that the higher the width of graphene micro-ribbon, the higher the total channel conductance. Second, it is worth mentioning that the variation in conductance values is greatest among the devices with 50 μm SLG width. This reverse conductance variability (X-axis) decreases as the SLG width increases to 100 μm and ultimately approaches zero for SLG width equal to 200 μm. The latter observation suggests that sensors based on SLG with IDE are less prone to graphene inhomogeneity for widths above 100 μm.

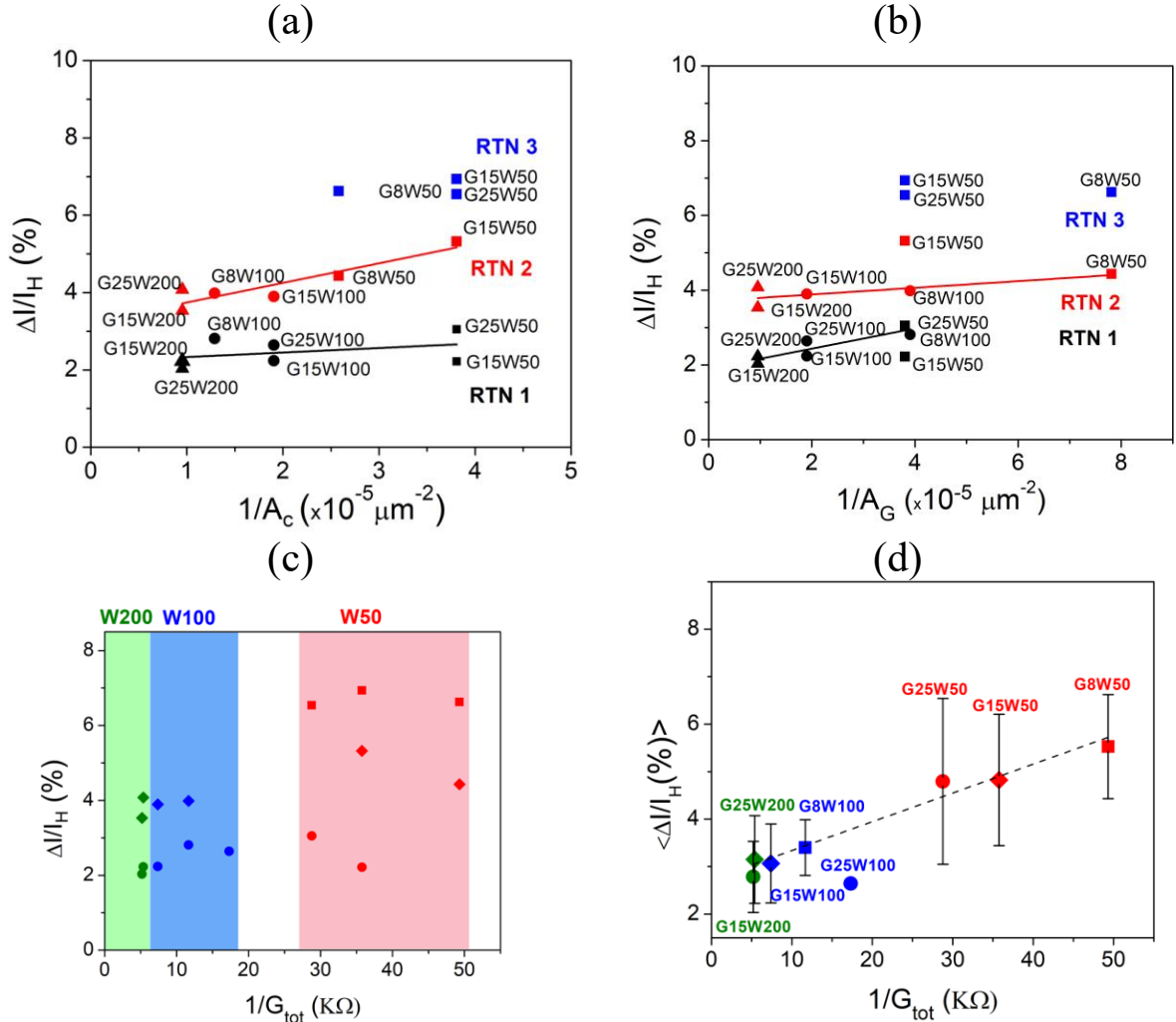

Figure 8: Plots of (a) $\Delta I/I_H$ (%) $-1/A_c$ was created to distinguish the impact of metal regions from that of free graphene in the detected RTN events among IDEs (b) $\Delta I/I_H$ (%) $-1/A_G$ for three different trap regions RTN-1, RTN-2, and RTN-3. (c) $\Delta I/I_H$ (%) $-1/G_{tot}$ reveals that high free graphene areas have the greatest amount of charge carriers. Moreover, the high variability in conductance values indicates that electron mobility is not the same as graphene area approaching the nanoscale. (d) $<\Delta I/I_H$ (%)$>$ $-1/G_{tot}$ suggests a clear correlation between conductance and the mean values of $\Delta I/I_H$ (%).

This conductance variability in devices of 50 μm SLG width probably originates from changes in electron mobility due to the limited number of current pathways (percolation effect). Another factor that may affect conductance is the increasing effect of SLG edges (armchair, zigzag) as the free graphene micro-ribbon size approaches the nanoscale. Finally, despite the spread of points in Fig. 8(c) and the absence of a clear linear correlation, Fig. 8(d) reveals a clear linear correlation ($R^2 = 0.92$) between the mean value of $\Delta I/I_H$ (%) from each device versus $1/G_{tot}$, regardless of the exact device geometry. Black vertical lines denote the standard deviation of $\Delta I/I_H$ (%) for each device. To conclude, we assess that this final plot demonstrates that even in cases of micro- or nano-scale graphene sheets, where both carrier density and mobility can vary a lot from region to region or from device to device, the total measured conductance can function as a metric for the expected amplitudes of RTN. This means that, in general, a device with higher conductance levels (not necessarily achieved through geometrical characteristics) would exhibit RTN of lower normalized amplitudes. This information is valuable for the reduction of the SNR in graphene sensor performance, or, conversely, can be used in noise-enhanced sensing [26] [27], where RTN is used as a sensing tool and needs to be high.

# IV. Conclusions

A thorough investigation has been conducted on IDE-GMR devices with different geometric topologies. Initially, I-V characteristics revealed Ohmic behavior for currents that do not exceed 20 μA and a self-heating effect for higher current values. Through statistical analysis of specific time-domain windows from noise signals, up to three discrete RTN levels were revealed, originating from different regions of the IDE-GMR. Specifically, it was proved that the generation of different RTN levels is due to traps lying either at the $SiO_2$/SLG or SLG/Al interfaces. Finally, the correlation between GMR width and electron density was investigated.

# Acknowledgements

This work was supported in part by the research project "LIMA-chip" (Proj. No. 2748) which is funded by the Hellenic Foundation of Research and Innovation (HFRI) respectively.